\documentclass[twocolumn,10pt,aps,longbibliography,floatfix]{revtex4-2}
\usepackage{amsmath,amssymb,amsfonts}
\usepackage{algorithmic}
\usepackage{graphicx}
\usepackage{textcomp}
\usepackage{float}
\usepackage[caption = false]{subfig}
\usepackage{enumitem}
\usepackage{multirow}
\usepackage{array}
\usepackage{bbm}
\newcolumntype{M}[1]{>{\centering\arraybackslash}m{#1}}
\RequirePackage{filecontents}
\setlist{nolistsep}
\usepackage[separate-uncertainty = true]{siunitx}
\def\BibTeX{{\rm B\kern-.05em{\sc i\kern-.025em b}\kern-.08em
    T\kern-.1667em\lower.7ex\hbox{E}\kern-.125emX}}
\PassOptionsToPackage{hyphens}{url}\usepackage{hyperref}

\makeatletter
\renewcommand*\env@matrix[1][*\c@MaxMatrixCols c]{%
  \hskip -\arraycolsep
  \let\@ifnextchar\new@ifnextchar
  \array{#1}}
\makeatother
\usepackage{cleveref}

\begin{document}

\title{Optimization of Tensor Network Codes with Reinforcement Learning}
\author{Caroline Mauron, Terry Farrelly and Thomas M. Stace}
\affiliation{ARC Centre for Engineered Quantum Systems, School of Mathematics and Physics, The University of Queensland, St Lucia, Queensland 4072, Australia}

\begin{abstract}

Tensor network codes enable structured construction and manipulation of stabilizer codes out of small seed codes. Here, we apply reinforcement learning to tensor network code geometries and demonstrate how optimal stabilizer codes can be found. Using the projective simulation framework, our reinforcement learning agent consistently finds the best possible codes given an environment and set of allowed actions, including for codes with more than one logical qubit. The agent also consistently outperforms a random search, for example finding an optimal code with a $10\%$ frequency after 1000 trials, vs a theoretical $0.16\%$ from random search, an improvement by a factor of 65. 

\end{abstract}

\maketitle

\section{Introduction}
\label{sec:introduction}

Developing efficient and reliable quantum error correction schemes is  critical for achieving large-scale quantum computing \cite{b6}. %While the no-cloning theorem prevents classical repetition codes, it is nevertheless possible to encode logical information into several physical qubits \cite{b7,b8}. 
In quantum error correction theory, the stabilizer formalism \cite{b7,b8,b5} has proven particularly useful in producing codes and understanding their structure. 

Building on the successful use of tensor networks in other areas of quantum information science \cite{b10, b11, b20, b21, b33, b34, b35}, tensor-network codes were recently introduced \cite{b2} as a method to construct new codes by applying tensor network operations to existing codes, including codes with multiple logical qubits \cite{b22}. For efficiently contractible tensor-network codes, this method comes with natural tensor-network decoders and efficient ways to compute the distribution of Pauli-weights of logical operators and stabilizers, allowing one to easily read off the code distance \cite{b3}. 

This capability opens up the possibility of using optimization strategies such as reinforcement learning to find  tensor-network codes that optimise some target properties. Various machine-learning techniques, including deep learning and reinforcement learning (RL), have already been widely applied in quantum error correction, mostly to decoding problems \cite{b23,b24,b25,b26,b27,b28,b29,b30,b31,b32}, but also recently as a tool to optimizing surface code layout \cite{b15}.

%\tf{Needs a good size paragraph or two explaining what we do in more detail.  For example, see how https://arxiv.org/pdf/1801.01879.pdf discusses what they do with some detail in the introduction.}

In this paper, we treat the search for good codes as a `game', in which the starting pieces of the game are a finite initial set of seed code tensors \cite{b2,b3}, which, when contracted together yield various possible stabilizer codes.  For our purposes, better codes are those that have high distance for a given number of physical qubits, and so there is a natural performance metric associated with a code.  The choice of contractions of seed tensor indices determines the nature of the resulting code.  Finding a good choice of contractions by brute force search is computationally expensive, and so this contraction-finding game is  a promising setting for reinforcement learning. 

We begin by reviewing the construction of stabilizer codes and of tensor-network codes in \cref{sec:tncodes}. We then describe in \cref{sec:rl} the specific reinforcement learning algorithm we employ in this paper, and present some applications and results in \cref{sec:apps}. We briefly conclude and suggest areas of further research in \cref{sec:conclusion}.

\section{Tensor-Network Codes}
\label{sec:tncodes}

Tensor-network codes are built on stabilizer codes, which are an important class of quantum error correcting (QEC) codes whose construction uses group theory.  We consider here an $\left[[n,k,d\right]]$ stabilizer quantum error-correcting code to encode  $k$ logical qubits into $n$ physical qubits, where the code distance $d$ is the Pauli-weight of the smallest logical operator.%with a distance $d$ such that errors with Pauli-weights less than d are detectable. 
% % i.e.\ those  for which . 

In the stabilizer formalism, we define the Pauli group $\mathcal{G}_n$ as the set of all operators $z \sigma^{i_1} \otimes \dots \otimes \sigma^{i_n}$, where $z \in \{ \pm 1, \pm i \}$, and $\sigma^{0} = \mathbbm{1}, \sigma^{1} = X, \sigma^{2}=Y, \sigma^{3}=Z$ are the single-qubit identity operator and three Pauli operators.

The subspace of the Hilbert space used to encode logical information, known as the codespace, is fully determined by an abelian group of operators $\mathcal{S} \subset \mathcal{G}_n$ called stabilizers: every state $|\psi\rangle$ in the codespace satisfies $S |\psi\rangle = |\psi\rangle$ for all stabilizers $S_i \in \mathcal{S}$. If $\mathcal{S}$ is generated by $r$ independent operators, the dimension of the codespace is $2^{n-r}$, corresponding to $k=n-r$ encoded logical qubits.

The logical operators of the code form a non-abelian group $\mathcal{L} \subset \mathcal{G}_n$ that is generated by $k$ $Z$-type and $k$ $X$-type logical operators (meaning they anticommute with each other). Logical operators, denoted respectively as $Z_\alpha$ and $X_{\beta}$, with $\alpha,\beta\in[1,...,k]$, commute with all stabilizers $S_i$.

%\tf{Can probably comment out this bit, and leave the pure errors out in the next paragraph.  Just explain the syndrome and that decoding is hard.}
The final important group of operators is the abelian group of pure errors $\mathcal{E} \subset \mathcal{G}_n$, which is generated by $n-k$ operators $E_j$ that satisfy $E_j S_i = (-1)^{\delta_{ij}} S_i E_j$. The entire Pauli group $\mathcal{G}_n$ is generated by all $S_i, Z_{\alpha}, X_{\beta}$ and $E_j$.

In order to detect errors, each stabilizer generator is measured, producing an error syndrome vector $\vec{s}$ of length $r$, with $s_i=0$ if the error commuted and $s_i=1$ if the error anticommuted with the stabilizer $S_i$. It is straightforward to find an error consistent with the measured syndrome, for example the operator $E(\vec{s}) = \prod_i E_i^{s_i}$, but the degeneracy of quantum codes implies that many different errors may give rise to the same syndrome, e.g., $E(\vec{s}) S$ for any $S \in \mathcal{S}$ also has the same syndrome as $E(\vec{s})$. Finding the \emph{optimal} correction operator given the syndrome is extremely difficult in general \cite{b12,b13}.

Among stabilizer codes, the five-qubit code is important because it is the smallest quantum error correcting code that can protect a logical qubit from any arbitrary single qubit error, and will be used further in our applications. The code generators and its logical operators are given in \cref{tab:qubcode5}(left). 
%\tf{Don't need next sentence:} We can easily confirm that all stabilizers anticommute with each other, all logical operators anti-commute with each other, and all stabilizers commute with all logical operators.
This $[[5,1,3]]$ code can also be naturally adapted to a six-qubit \emph{stabilizer state} $[[6,0]]$ by adding a physical qubit on which the original code stabilizers have no action, but the logical operators are augmented by a non-trivial Pauli operator, as displayed in \cref{tab:qubcode5}(right).  

Because the five-qubit code is the most elementary quantum error correcting code, we primarily use copies of it, and the related six-qubit stabiliser state, as the seed tensors from which to create more complex codes.   To demonstrate the flexibility of our learning scheme, we also include  the $[[10,1,4]]$ code tensor, which is the smallest quantum code with distance 4.  The stabilizer matrix for this code is obtained from \cite{b14}, and we follow the procedure given in \cite{b1} to obtain the logical operators.

\begin{table}[t]
\centering
\setlength\extrarowheight{5pt}

\begin{tabular}{|M{1.2cm}|M{0.3cm}M{0.3cm}M{0.3cm}M{0.3cm}M{0.3cm}|}
\hline
Qubit &  1 &  2 & 3 & 4 & 5 \\
\hline
$S_1$ &  $X$ &  $Z$ & $Z$ & $X$ & $\mathbbm{1}$ \\
$S_2$ &  $\mathbbm{1}$&  $X$ & $Z$ & $Z$ & $X$  \\
$S_3$ &  $X$ & $\mathbbm{1}$ & $X$ & $Z$ & $Z$  \\
$S_4$ &  $Z$ & $X$ & $\mathbbm{1}$ & $X$ & $Z$   \\
\hline
$X_1$ &  $X$ & $X$  & $X$ & $X$ & $X$   \\
$Z_1$ &  $Z$ & $Z$  & $Z$ & $Z$ & $Z$   \\
\hline
\end{tabular}
\quad
\begin{tabular}{|M{1.2cm}|M{0.3cm}M{0.3cm}M{0.3cm}M{0.3cm}M{0.3cm}M{0.3cm}|}
\hline
Qubit &  1 &  2 & 3 & 4 & 5 & 6\\
\hline
$S_1$ &  $X$ &  $Z$ & $Z$ & $X$ & $\mathbbm{1}$ & $\mathbbm{1}$ \\
$S_2$ &  $\mathbbm{1}$&  $X$ & $Z$ & $Z$ & $X$ & $\mathbbm{1}$ \\
$S_3$ &  $X$ & $\mathbbm{1}$ & $X$ & $Z$ & $Z$ & $\mathbbm{1}$  \\
$S_4$ &  $Z$ & $X$ & $\mathbbm{1}$ & $X$ & $Z$ & $\mathbbm{1}$  \\
\hline
$S_5$ &  $X$ & $X$  & $X$ & $X$ & $X$ & $X$   \\
$S_6$ &  $Z$ & $Z$  & $Z$ & $Z$ & $Z$ & $Z$   \\
\hline
\end{tabular}
\caption{(left) Stabilizer generators and logical operators for the five-qubit code $[[5,1,3]]$, and (right) for the six-qubit stabilizer state (i.e., purified five-qubit) $[[6,0,3]]$. (Note the distance is trivially zero if defined instead as the Pauli-weight of the smallest logical operator)}
\label{tab:qubcode5}
\end{table}

Following Ref.\ \cite{b2},  we use tensors to describe and build stabilizer codes. Operators are represented by tuples of integers, for example the first stabilizer generator of the $[[5,1,3]]$ code %\tf{maybe use the 5 qubit code because it's shorter and the table of stabs is right above} 
$S_1 = XZZ X \mathbbm{1} = \sigma^1 \sigma^3 \sigma^3 \sigma^1 \sigma^0 $ is represented by $(1, 3, 3, 1, 0)$. Then for each logical operator $L \in \mathcal{L}$, we define the rank-$n$ tensor
\begin{equation}
T(L)_{g_1, \dots , g_n} = 
\begin{cases}
1 &\text{if } \sigma^{g_1} \otimes \dots \otimes \sigma^{g_n} \in \mathcal{S} L \\
0 &\text{otherwise,}
\end{cases}
\end{equation}
where $g_j \in \{0,1,2,3\}$ and $\mathcal{S} L$ is the set of all operators of the form $SL$ with $S \in \mathcal{S}$. For example, $T(\mathbbm{1})_{g_1, \dots, g_n}$ describes the stabilizer group and $T(X_1)_{g_1, \dots, g_n}$ describes the coset corresponding to logical operator $X_1$. 

\begin{figure}[t]
\includegraphics[width = 3.0in]{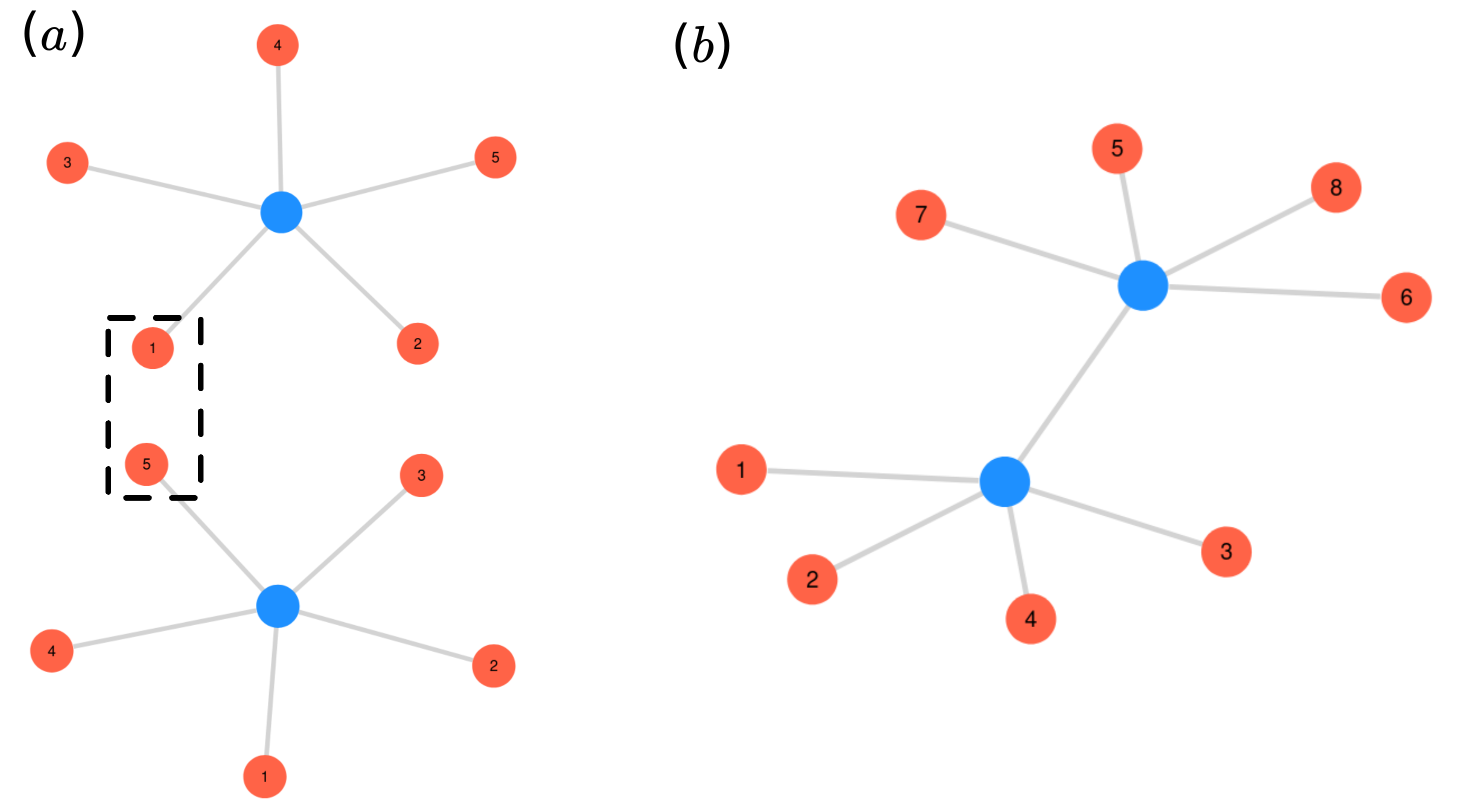}
\caption{Example of contracting tensor-network codes.  (a) The blue vertices represent the two five-qubit code tensors; the red vertices represent qubits, labelled 1 to 5 corresponding to the qubit labelling in \cref{tab:qubcode5}.  The dashed box indicates the two legs we want to contract to form a bigger code.  (b) The result of contracting the two legs.  In this case we have a [[8,2,3]] error correcting code.  Tensor network codes were simulated using the TensorNetworkCodes.jl package \cite{b3,TensorNetworkCodes.jl}.}
\label{fig:tn_contract}
\end{figure}

New quantum error-correcting codes can then be created from tensor-network contractions, referred to later as `tensor fusion'.  %  , trace (hereforth referred to as combination and fusion respectively) [IS THIS CORRECT?] and contraction 
Tensor fusion is equivalent to measuring $XX$ and $ZZ$ operators on the physical qubits corresponding to the contracted tensor legs.  Tensor fusion creates a new tensor network code as long as a technical condition proved in \cite{b2} is satisfied, which is equivalent to the $XX$ and $ZZ$ measurements not measuring any logical degree of freedom.  An example of such an operation is illustrated in \cref{fig:tn_contract}, in which two copies of the five-qubit code are fused at a specific pair of nodes, to form a new [[8,2,3]] code.
%Graphical representation of the sample codes discussed in the previous section is provided in \cref{fig:samplecodes}. 

% \begin{figure}[h]
% \subfloat[${[[}5,1,3{]]}$]{\includegraphics[width = 1.1in]{code513.png}} 
% \subfloat[${[[}6,0,0{]]}$]{\includegraphics[width = 1.1in]{code600.png}} 
% \subfloat[${[[}10,1,4{]]}$]{\includegraphics[width = 1.1in]{code1014.png} } 
% \caption{Three codes are provided here in graph representation for illustration. Logical tensor nodes are colored in blue (central nodes on the left and right chart), tensors with no logicals are in green (central node on the middle plot), and physical qubits are in red.}
% \label{fig:samplecodes}
% \end{figure}

\section{Reinforcement Learning}
\label{sec:rl}

It was first suggested in \cite{b2} that a reinforcement learning (RL) strategy may be used to search for good error-correcting codes among codes with tensor-network geometries that are efficiently contractible. We build here on recent successful use of RL algorithms to optimize surface codes \cite{b15} using a projective simulation model for RL. Projective simulation is a physics-motivated framework for artificial intelligence \cite{b16}. As in generic RL settings, an agent takes actions within its environment, receives rewards, and learns to optimize its actions based on these rewards.

\subsection{Environment}
\label{subsec:env}

For our code optimisation, the environment is initialized with several copies of one or more seed tensor-network codes as defined in \cref{sec:tncodes}. In order to use this environment in an RL setting, we must define what constitutes a state (known as ``percept'', in the PS framework) describing the environment, an action (or move) of the agent within this environment, and a reward policy that provides feedback to the agent.

We represent the state of the environment (known as percept in the projective simulation framework) by the stabilizer generators of the QEC code as defined in \cref{sec:tncodes}. 

\begin{figure}[t]
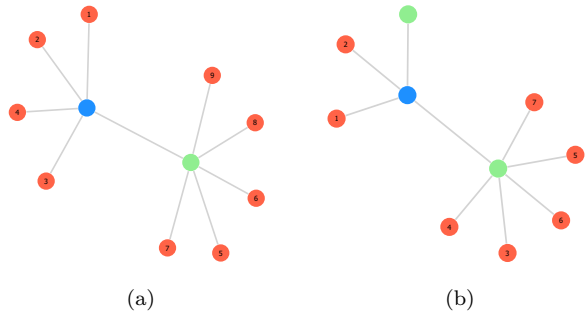

\subfloat[]{\includegraphics[width = 1.5in]{code_fusion01.png} \label{subfig:fusion01}}
\quad
\subfloat[]{\includegraphics[width = 1.5in]{code_fusion00.png} \label{subfig:fusion00}} 
\caption{The fusion of one five-qubit node with one six-qubit stabilizer state is depicted on \cref{subfig:fusion01}. On the right, \cref{subfig:fusion00} shows a fusion of the five-qubit node with itself, which is equivalent to a contraction of two physical qubits on that node with a Bell pair stabilized by XX and ZZ (shown as a new green node)}
\label{fig:rlactions}
\end{figure}

An agent can then move within its environment by sequentially performing tensor-network operations for a pre-determined number of steps $s$. Specifically, we narrowly define an action here by only specifying two tensor nodes (which need not be distinct) in the tensor network, and performing a fusion between two arbitrary physical qubits attached to these nodes. Some examples of possible actions are depicted in \cref{fig:rlactions}. 

Actions are symmetric with respect to nodes, so that in a tensor-network codes with $n$ nodes, there are $n(n+1)/2$ possible actions. For an environment with $n$ nodes that allows $s$ steps, the total number of possible sequences is therefore given by 
\begin{equation}
N= { s-1 + {n(n+1)}/{2}  \choose s}.
\label{eq:Nposs}
\end{equation} 
Some of these sequences will however not be possible, for example those that involve fusing a node whose physical qubits have all been fused previously. The list of allowed actions is made available to the agent at every step. (This is not strictly necessary as the environment could instead have issued a negative reward and let the agent learn from it, at the cost of slower learning).

Finally, a key component of any RL framework is to define the rewards provided by the environment to the agent. It is in general not straightforward to compare QEC codes and determine which is superior to another, potentially requiring computationally-heavy error simulations. Here, we choose to focus on the distance of the stabilizer code, in the sense that for a given number of physical qubits, higher distance is better. The choice made above to only allow simple fusion actions (as opposed to more complex combinations or contractions) also simplifies the choice of a reward policy. 

Since every action decreases the number of physical qubits by 2, higher distance is always better, and defining the reward as the change in code distance appears to be a sensible choice. One nuance is that a fusion of two qubits may lead to a reduction in distance but still produce an optimal code, in the sense that there does not exist any stabilizer code with a larger distance for this number of physical and logical qubits. In this case, the environment returns a reward of +1, instead of the negative reward it otherwise would have. (The optimal distances per number of physical and logical qubits in stabilizer codes up to $n=50$ is known from numerical calculations, and listed in \cite{b14}.) 
In cases of tensor network codes with more than one logical qubit, there exists some edge cases of actions that are not in the list of obviously-impossible actions, but nevertheless are not possible. In such cases, the environment returns a negative reward equal to the previous distance and terminates this sequence.

\subsection{Agent}

The core component of the agent is its two-layered network, represented by a directed, bipartite weighted graph, where the two disjoint sets comprise the percepts $P$ and actions $A$, respectively. In this network, each percept (state) $s_i \in P$ is connected to an action $a_j \in A$ via a directed edge $(i, j)$ which represents the possibility of taking action $a_j$ given the percept $s_i$ with probability $p_{ij} = p(a_j | s_i)$.
A time-dependent weight called $h$-value, $h^{(t)}_{ij}$ is associated with each edge $(i,j)$ and is updated at each time step via the matrix formula
\begin{equation}
  \begin{aligned}
h^{(t+1)} &= h^{(t)} + \lambda^{(t)} g^{(t)} + \gamma (1 - h^{(t)}) \\
g^{(t+1)} &= (1 - \eta) g^{(t)}
  \end{aligned}
\end{equation}
where
\begin{itemize}
\item $\lambda^{(t)}$ is the reward provided by the environment
\item $g$ is the so-called glow matrix, containing memory of past actions taken in that trial (also known as episode)
\item $\eta$ is the glow parameter, determining how quickly past experiences are forgotten within one trial ($\eta=1$ means the agent considers each reward to be a direct consequence of its last action only, while $\eta=0$ considers it a consequence of the entire sequence of actions with equal weights on each action).
\item $\gamma$ is a damping parameter that is most useful in case of environments that are changing over time, allowing the agent to partially forget what it has learnt.
\end{itemize}

The transition probability from percept $s_i$ to action $a_j$ is calculated via the softmax function \cite{b18}
\begin{equation}
  \begin{aligned}
p^{(t)}_{ij} &= \frac{e^{\beta h^{(t)}_{ij}}} { \sum_{k} e^{\beta h^{(t)}_{ik}} }
  \end{aligned}
\end{equation}
where the sum is taken over all possible actions $a_k \in A$ and $\beta > 0$ is the softmax parameter.
Finally the transition probabilities are re-normalized to only allow viable actions from each percept as determined by the environment.

\subsection{Interaction}

The environment is initialized with a large tensor network code and the agent aims to find the code with the highest possible distance that can be built from performing $s$ fusions of physical qubits on that starting code. Our agent learns by performing multiple trials (stochastic processes), with each trial consisting of $s$ steps. In order to assess the performance of an agent, we typically do 20 simulations (each with a different seed) of 1000 trials. 

At each step, the agent decides on an action by drawing a random sample of the full list of actions using the transition probabilities it calculated. The environment then performs the corresponding fusion, calculates the reward, its new state (the stabilizers of the new code) and the list of actions that are possible from that state. It passes that information on to the agent, which updates the $h$ and $g$ matrices and (unless that episode is terminating) decides on the next action. The agent does not know in advance the full set of potential percepts, and instead continuously updates its set of percepts as it encounters them. The $g$-matrix is reset to 0 at the beginning of each trial, while the $h$-matrix is reset to 1 at the beginning of each simulation only.

\section{Applications}
\label{sec:apps}

Our simulations code was developed in Python, and used both the TensorNetworkCodes.jl Julia package \cite{b3,TensorNetworkCodes.jl} and the ProjectiveSimulation Python package \cite{b19}. Taking inspiration from the example showing how to generate the $[[13,1,5]]$ code with tensor networks in Ref \cite{b3}, we initialize our environment with a code built from combining one five-qubit code, and three six-qubit stabilizer states, which, before any fusion operations are performed, trivially corresponds to a $[[23,1,3]]$ code. We then ask the RL agent to find the best code that can be obtained after a sequence of $s= 1$ to 6 fusions (using learning hyper-parameters $\beta=1, \eta=0.05, \gamma=0$). To establish the learning performance the RL, we also compute all possibilities directly with a brute force search to obtain the definitively best code after $s$ steps. Results are displayed in \cref{tab:code3step5}.

\begin{table}[t]
\centering
\setlength\extrarowheight{3pt}
\begin{tabular}{M{1.2cm}|M{3.0cm}|M{3.0cm}}
Steps, $s$ &  RL best code & Brute force search \\
\hline
1 & $[[21,1,3]]$ & $[[21,1,3]]$ \\
2 & $[[19,1,3]]$ & $[[19,1,3]]$ \\
3 & $[[17,1,5]]$ & $[[17,1,5]]$ \\
4 & $[[15,1,5]]$ & $[[15,1,5]]$ \\
5 & $[[13,1,5]]$ & $[[13,1,5]]$ \\
6 & $[[11,1,3]]$ & $[[11,1,3]]$ \\
\end{tabular}
\caption{Codes with the maximum distance obtained after $s$ fusion steps starting from an initial $[[23,1,3]]$ code built by combining one five-qubit code and three six-qubit stabilizer states.}
\label{tab:code3step5}
\end{table}

We see that the optimal code found by the RL algorithm agrees with the brute force search optimum. We note that after 3 and 4 steps, it finds codes that are optimal in the sense that they have the maximum distance for that number of physical qubits \cite{b14}. After 6 steps, it does not find the optimal  $[[11,1,5]]$ code, but a systematic search of all possibilities verified that this code was not reachable at all via this method.  

\begin{figure}[b]
\subfloat[Code A graph]{\includegraphics[width = 1.5in]{codeA.png} \label{subfig:codeA}}
\quad
\subfloat[Code B graph]{\includegraphics[width = 1.5in]{codeB.png} \label{subfig:codeB}}  \\
%\subfloat[]{\includegraphics[width = 2.0in]{hist1.png}} \\
\subfloat[Operator weights comparison for Code A and Code B.]{\includegraphics[width = \columnwidth]{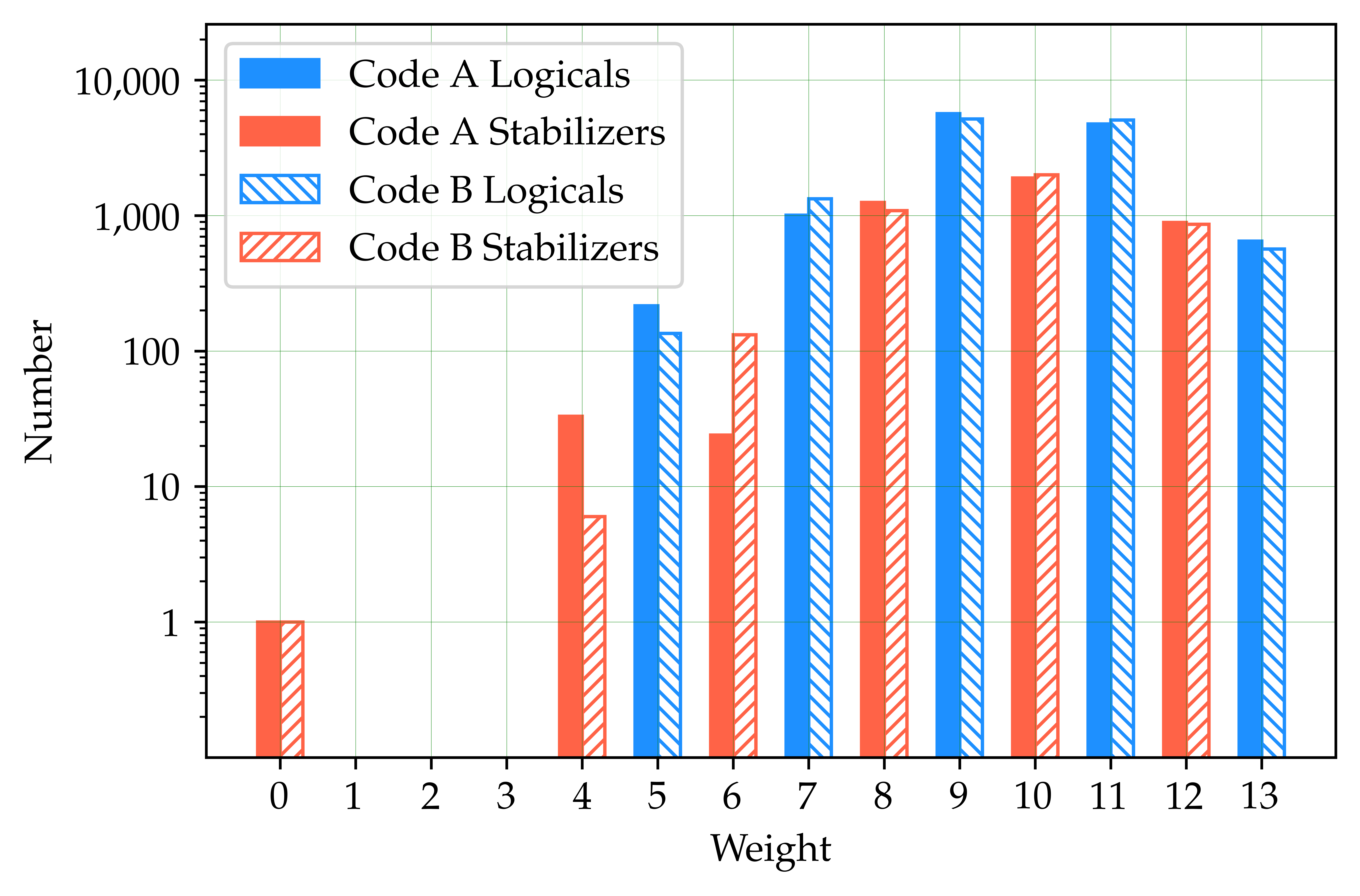} \label{subfig:doublehist}} \\
%\subfloat[]{} \\
\caption{Graph representations and operator weights histogram of two different $[[13,1,5]]$ codes found by the RL algorithm after starting from a $[[23,1,3]]$ code and optimizing a sequence of 5 fusions. We note the codes are not equivalent, having different operator weights. In this case, the lower logical weight for code B suggests it is likely to be a better code, since there are fewer undetectable low-weight errors.}
\label{fig:codecomp}
\end{figure}

One notable finding here is that RL can find inequivalent codes with the same signature $[[n,k,d]]$. For example, two codes that were obtained after $s=5$, different runs of the RL search finds distinct $[[13,1,5]]$ codes, shown as `Code A' and `Code B' in  \cref{fig:codecomp}.  We establish that these are distinct by comparing the histogram of logical operator weights for the two codes. We display the graph representation of the codes and their operator weights histograms in \cref{fig:codecomp}.  We see that the smallest weight logical operator for both codes is $d=5$, but that there are different numbers of such minimum-weight operators.

Further, we find that RL can outperform a greedy search. For example, we see that that while $[[17,1,5]]$ is the optimal code found after 3 steps, the RL strategy is able to find a $[[13,1,5]]$ code after 5 steps that cannot be built from this $[[17,1,5]]$ code; that is, the sequence of actions to reach the $[[13,1,5]]$ code does not generate the $[[17,1,5]]$ as an intermediate step. Indeed, searching all codes that can be built from the $[[17,1,5]]$, one finds the code displayed on \cref{subfig:codeB} but not the code displayed on \cref{subfig:codeA}. This demonstrates that the algorithm is not necessarily greedy, potentially improving its performance and usefulness on larger codes.

\begin{figure}[b]
\includegraphics[width = \columnwidth]{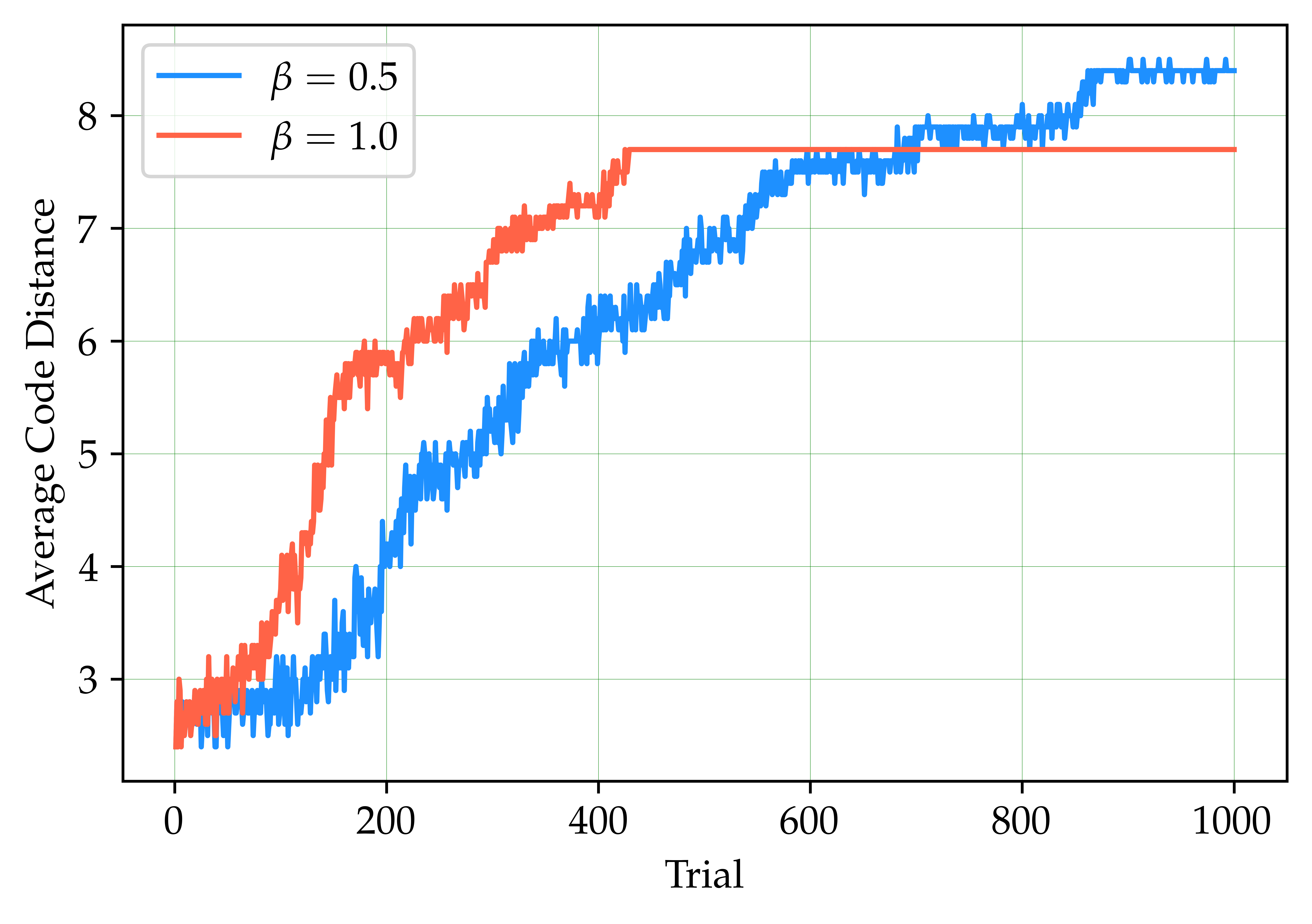}
\caption{Average code distance as the RL agent progressively finds the optimal code. This example shows the case of an initial $[[35,1,3]]$ code and $s=5$ fusions in order to ultimately find the $[[25,1,9]]$ code. We see the role of the softmax parameter $\beta$, one of the parameters controlling the tradeoff between exploration and exploitation. With a higher $\beta$, the agent settles on a code quicker after receiving positive rewards, but this may be at the expense of never finding the better code. With $\beta=0.5$, 16 out of our 20 simulations return the $[[25,1,9]]$ code after 1000 trials, against only 11 with $\beta=1.0$}
\label{fig:avgdisttrial1}
\end{figure}

In order to assess the performance of our RL algorithm, we now apply it to codes that are still small enough to also be found via brute force search, but large enough that the probability of quickly finding the optimal codes by random search is now quite small. We repeat the simulations of the previous subsection with initial codes that include one five-qubit code and either 4 or 5 six-qubit stabilizer states, which correspond to uncontracted $[[29,1,3]]$ and $[[35,1,3]]$ respectively.  Outputs are summarized in \cref{tab:outputs}. In all cases where a brute-force search was done, RL found the best-possible code, including several optimal codes, in the sense of the highest-possible distance for a set number of physical qubits.  

\begin{table}[t!]
\centering
\setlength\extrarowheight{3pt}
\begin{tabular}{M{2.0cm}|M{2.0cm}|M{2.0cm}}
Initial Code & $[[29,1,3]]$ & $[[35,1,3]]$ \\
\hline
Steps & \multicolumn{2}{M{4.0cm}}{RL best code} \\
\hline
1 & $[[27,1,3]]$ & $[[33,1,3]]$ \\
2 & $[[25,1,3]]$ & $[[31,1,3]]$ \\
3 & $[[23,1,5]]$ & $[[29,1,5]]$ \\
4 & $[[21,1,7]]^*$ & $[[27,1,7]]$ \\
5 & $[[19,1,5]]$ & $[[25,1,9]]^*$ \\
6 & $[[17,1,5]]$ & $[[23,1,7]]$ \\
7 & $[[15,1,5]]^*$ & $[[21,1,7]]^*$ \\
8 & $[[13,1,5]]^*$ & $[[19,1,7]]^*$ \\
9 & $[[11,1,3]]^{\dagger}$ & $[[17,1,5]]^{\dagger}$ \\
10 & $[[9,1,3]]^*$ & $[[15,1,5]]^*$ \\
11 & $[[7,1,3]]^*$ & $[[13,1,5]]^*$ \\
\end{tabular}
\caption{Code with the maximum distance obtained after $s$ fusion steps starting from an initial $[[29,1,3]]$ (resp. $[[35,1,3]]$) code built by combining one five-qubit code and four (resp. five) six-qubit stabilizer states. Codes with a $*$ are optimal in the sense that their distance is the maximum reachable with this number of physical qubits \cite{b14}. In the cases marked with a $\dagger$, we were not able to run the brute force search. All other codes were verified by a brute force search to be the maximum distance that could be reached from that initial code. }
\label{tab:outputs}
\end{table}

An example of the RL progression showing how the average code distance evolves over time (as the algorithm learns from more trials) is displayed in \cref{fig:avgdisttrial1}. It illustrates the typical reinforcement learning trade-off between exploration and exploitation. A higher softmax parameter $\beta$ allows the RL agent to reach a higher distance faster, but reduces the probability of reaching the best possible code, as it may cause the agent to get stuck on a non-optimal code.

\begin{figure}[t]
\includegraphics[width = \columnwidth]{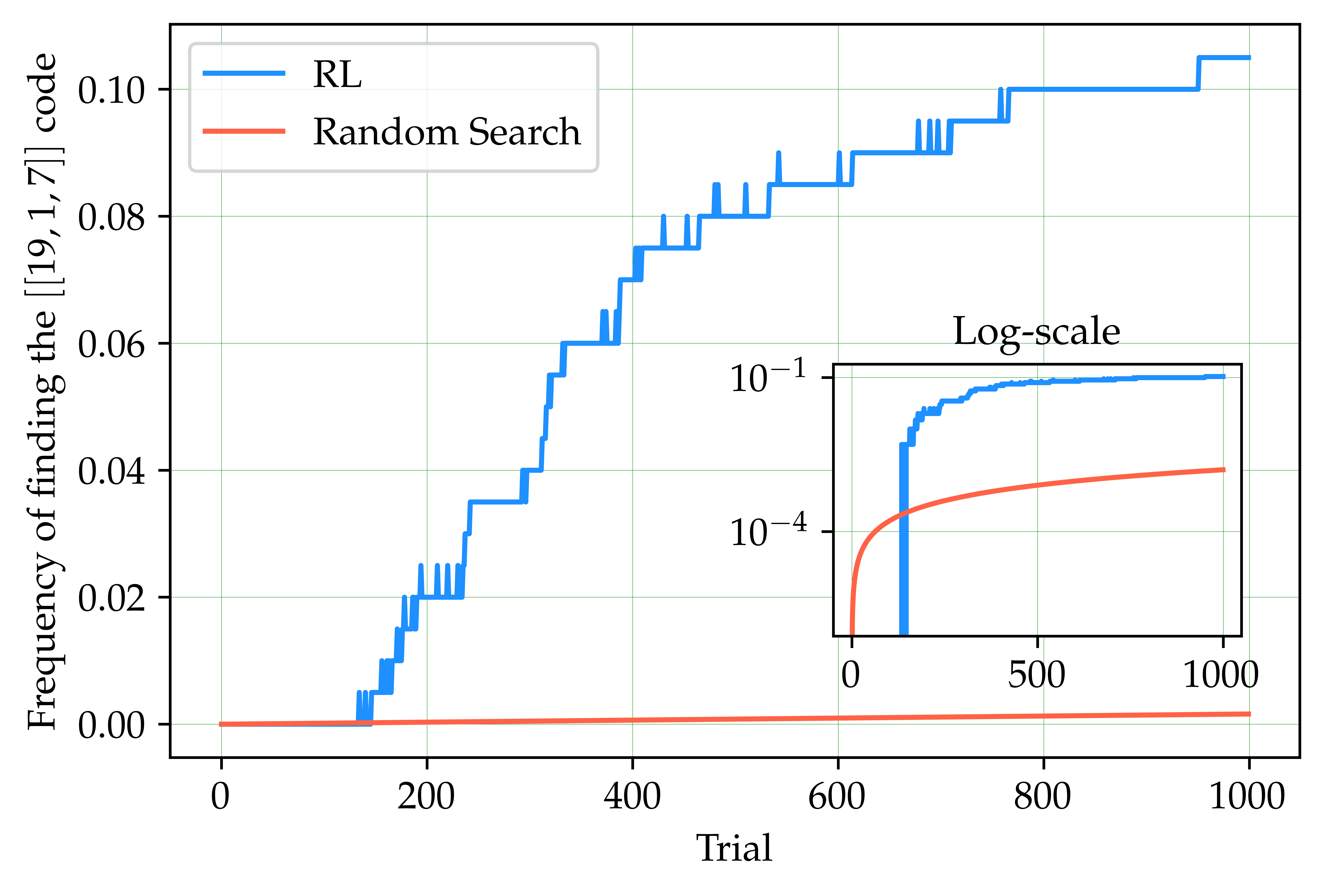} 
\caption{
Probability of finding the $[[19,1,7]]$ over 1000 trials using the RL agent compared to a random search, when starting with a $[[35,1,3]]$ code and allowing 8 fusions. For the random search probability, a brute force search was first conducted showing only 5 out of the N=$3,108,105$ possible fusion sequences lead to the maximum-distance $[[19,1,7]]$ code. The theoretical random search probability for given trials was then calculated using \cref{eq:bernoulliproba}. For the RL probability, we ran 200 simulations (each with a different random seed), and the probability is returned as the frequency of the environment state being in a $[[19,1,7]]$ code at each trial. After 2000 trials, the random search probability is $0.16\%$, while the RL frequency is 10.2\%. }
\label{fig:proba}
\end{figure}

We now benchmark the performance of the reinforcement learning algorithm to random search, by comparing the success rate of RL against the likelihood of finding optimal codes with random search. Specifically, if we are able to compute the distance for all $N$ possibilities (from \cref{eq:Nposs}), and we find the code with the optimal distance $n$ times, the probability of having found this code after $t$ random search trials will be
\begin{equation}
p = 1 - \left(1- \frac{n}{N} \right)^t,
\label{eq:bernoulliproba}
\end{equation}
which can then be compared to the RL performance. A plot of the comparison over 2000 trials is displayed in \cref{fig:proba} for one example, where we see that the RL algorithm outperforms random search by a factor of 65.

\begin{figure}[h]
\includegraphics[width = \columnwidth]{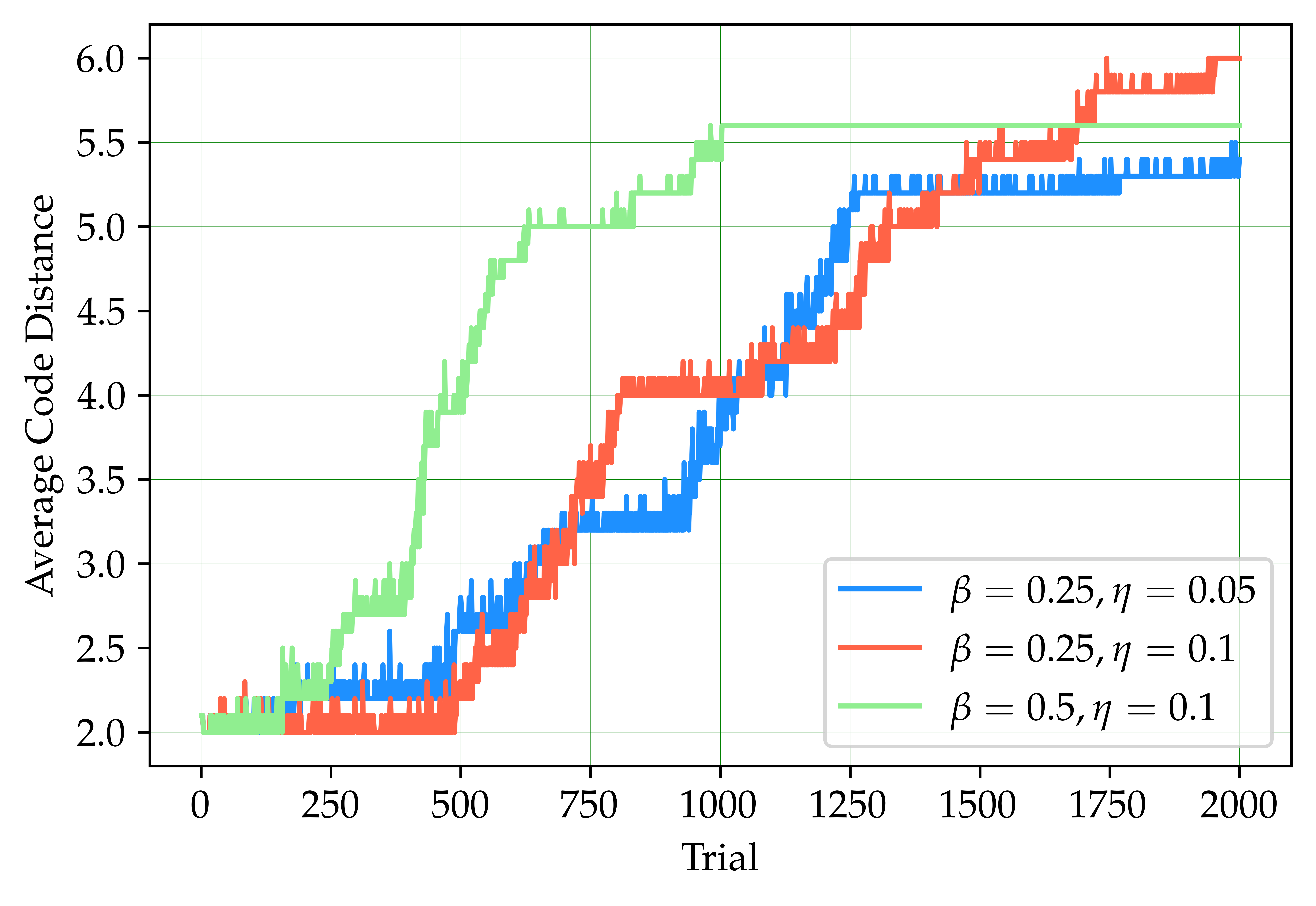}
\caption{Distance optimization by the RL agent in an environment with two logical qubits. This example shows the case of an initial $[[28,2,2]]$ code and 4 fusions in order to ultimately find the optimal $[[20,2,6]]$ code, under different RL parameters. With $\beta=0.25$ and $\eta=1$, the RL agent has settled on the $[[20,2,6]]$ code in all 20 simulations after 1000 trials.}
\label{fig:avgdisttrial}
\end{figure} 

We also applied our framework successfully to find codes with more than one logical qubit, specifically for $k=2$. In our first example, we show that RL can combine various seed tensors by starting with a code made of one $[[10,1,4]]$ code, one five-qubit $[[5,1,3]]$ code and one six-qubit $[[6,0]]$ stabilizer state, giving us an initial $[[21,2,3]]$ code. After 4 steps, our agent is able to find the $[[13,2,4]]$ code (the maximum-distance $k=2$ code with 13 physical qubits \cite{b14}) in 8 out of the 20 simulations. 

We can use this approach to ``boost" an error-detecting code into an optimal error-correcting code.  To show this, we start with one $[[4,2,2]]$ error-detecting seed code tensor and four six-qubit $[[6,0]]$ stabilizer states. Again, the reinforcement learning agent is in most cases able to find a (near) optimal code, which is the $[[20,2,6]]$ code \cite{b14}, after 4 steps. The evolution of the average distance under the RL framework in this case is displayed in \cref{fig:avgdisttrial}, also illustrating the possibility of fine-tuning the hyper-parameters.

\section{Conclusion}
\label{sec:conclusion}

Our main result is a demonstration that reinforcement learning can be employed to successfully look for quantum error-correcting codes with maximum distance, making use of tensor-network based methods to construct efficiently-contractible codes and calculate their distances. Using the projective simulation framework, the reinforcement learning agent can find the best possible code in its environment, and can find it with a much higher frequency than a random-search method. The agent was also demonstrated to find codes with the maximum possible distance in cases with more than one logical qubit.

There are many open avenues of research. The ability to also calculate the full distribution of operator weights via the same method opens the door to further optimize the codes towards desired properties. It may be interesting to introduce further restrictions on the allowed actions in the environment to reflect locality of a typical physical setup. Another avenue of research may be motivated by the observation that the agent was consistently unable to find some optimal codes such as $[[11,1,5]]$ and $[[16,2,6]]$. It would be interesting to establish whether these codes are `atomic' under the particular construction method considered here, in the sense that they cannot be constructed from contractions of other seed tensors.

\section*{Note added}
A few days prior to the completion of this paper, a related work appeared on the arXiv \cite{b17}, where the authors also look at using reinforcement learning to discover optimal quantum error correction codes using the `quantum lego' framework.

\begin{acknowledgments}
Numerical simulations were performed on The University of Queensland's School of Mathematics and Physics Core Computing Facility ``getafix''.
\end{acknowledgments}

~

%\bibliography{bib2}


\end{document}